\documentstyle[12pt,psfig]{article} 
\begin{document} 
% \draft command makes pacs numbers print %\draft
%\preprint{} 
\baselineskip18pt 
%\twocolumn 
\topskip=3cm
\begin{center}
{\Large\bf Possible Production of High-Energy Gamma Rays from Proton 
Acceleration in the Extragalactic Radio Source Markarian~501 }\\
% repeat the \author\address pair as needed 
\vskip1cm
K. Mannheim \\
Universit\"ats-Sternwarte,
Geismarlandstra{\ss}e 11,
G\"ottingen D-37083, Germany (kmannhe@uni-sw.gwdg.de)\\
\vskip0.5cm
{\it SCIENCE {\bf 279}, 684 (1998)}
\end{center}
%\date{\today} 
%\maketitle 
\newpage
%\begin{abstract} 
\newpage
\topskip=5cm
\centerline{\bf\sc ABSTRACT}
\vskip0.5cm

\noindent
%Gamma rays with an energy of 10~tera electron volts were 
%discovered from the active galaxy Markarian~501 with air-Cerenkov
%telescopes.
%
The active galaxy Markarian~501 was discovered with air-Cerenkov
telescopes at photon energies of 10~tera-electron volts.  
Such high energies may indicate
that the $\gamma$ rays from Markarian~501 are due to
the acceleration of protons rather than electrons.  Furthermore,
the observed absence of $\gamma$ ray 
attenuation  due to electron-positron
pair production in collisions with cosmic infrared photons
implies a limit of 2 to 4
nanowatt per squaremeter per steradian for the energy flux of
an extragalactic infrared radiation background at a
wavelength of 25 micrometers.
This limit provides important clues on the epoch of
galaxy formation.
\newpage
\topskip=0cm

Gamma rays ($\gamma$ rays) from cosmic sources impinging on Earth's atmosphere
initiate electromagnetic showers in which the energy of the primary $\gamma$ ray
is imparted among secondary electron-positron pairs.  The blue
Cerenkov light emitted by the pairs in the atmosphere can be
detected from the ground with optical telescopes triggering on the short
($\sim 1$~ns) optical pulses.  The technique has advanced considerably
in recent years ({\it 1}), and
some surprising discoveries have been made.  Among them is the detection of
the blazar Markarian~501 (Mrk~501) at energies above
10~TeV (1~TeV~=~$10^{12}$~eV) ({\it 2}).\\

Blazars are remote but very powerful sources characterized
by their variable polarized synchrotron emission. They are associated
with radio jets (bipolar outflows) emerging from
giant elliptical galaxies seen at small angles with the line of sight.
Mrk~501 is
$\sim 3\times 10^8$ light years from Earth but nevertheless
produces a tera-electron volt
$\gamma$ ray flux during outbursts that is many times stronger than that
of the Crab nebula, a supernova remnant inside our Milky Way at a distance of
only $6\times 10^3$~light years.  The radiation mechanism responsible for the
$\gamma$ rays could be either inverse-Compton scattering of low-energy photons
by accelerated electrons ({\it 3})
or pion production by accelerated protons.  In the
latter case, the sources could be among the long-sought sources of cosmic rays,
that is the
isotropic flux of relativistic particles with differential number
density ($N$)
spectrum $dN/dE\propto E^{-2.7}$ (for energies $E<10^3$~TeV)
mainly consisting of protons and ions ({\it 4}).\\

Particle acceleration in astrophysics is typically observed to be
associated with
(collisionless) shock waves when a supersonic flow of magnetized material hits a
surrounding medium.  Examples of shock waves are shell-type supernova remnants
(explosion of a massive star), plerions (pulsar wind), $\gamma$ ray
bursts (relativistic ejecta from the collapse of a compact stellar
object), or the jets ejected
from active galactic nuclei (collimated relativistic wind from the
accretion disk around a supermassive black hole).\\

In the theoretical picture of shock acceleration,
relativistic particles (protons, ions, electrons)
scatter elastically off turbulent fluctuations
in the magnetic field on both sides of the shock and thereby gain
energy because of the convergence of the scattering centers (approaching walls).
The acceleration time scale for the process can be written as
$t_{\rm acc}=\xi
r_{\rm g}c/u^2$ where $u$ denotes the velocity of the shock wave
(c is the speed of light)
and $r_{\rm g}\propto E/B$ denotes the radius of gyration 
of a particle with energy $E$
in a magnetic field of strength $B$.  
The effects of  shock obliquity, turbulence spectrum, and other unknowns
are conveniently hidden in an
empirical factor $\xi\ge 1$.
The most rapid (gyro-time scale) particle acceleration for
relativistic shocks corresponds to $\xi=1$ ({\it 5}).
Balancing the acceleration time scale with the energy loss
time scale due to synchrotron
radiation $t_{\rm syn}\propto B^{-2} E^{-1}$
one obtains the maximum
energy of the electrons $E_{\rm max}=10~(\xi/10)^{-0.5}(B/3\mu{\rm
G})^{-0.5}(u/10^8~\rm cm~s^{-1})$~TeV ({\it 6}).  The
observed 10~TeV $\gamma$ rays from the Crab nebula ({\it 7})
and the observed synchrotron x-rays
in shell-type supernova remnants ({\it 8}) (corresponding to 10~TeV
electrons) require $\xi\sim 1$~to~$10$.
Because protons lose less energy, they can reach
larger $E_{\rm max}$'s than electrons and give rise to $\gamma$ ray emission
even above $\sim 10$~TeV via pion production and subsequent pion decay.  
Although
shock acceleration theory predicts that most of the cosmic rays are
accelerated in supernova remnants ({\it 4}),
no definitive $\gamma$ ray signature has
yet been discovered.\\

It is commonly argued that the assumption of
electron acceleration also suffices to explain
the $\gamma$ rays from blazar jets such as Mrk~501 ({\it 9}).
Estimates of the magnetic field strength in the 
$\gamma$ ray emitting part of the jet in Mrk~501 then yield
values in the range
$B\sim0.04$ to $0.7$~G. 
This magnetic field is much
stronger than the one in supernova remnants and the associated
stronger cooling of the relativistic electrons due to
synchrotron energy losses reduces
$E_{\rm max}$ accordingly. 
The effect is almost compensated by
the high shock wave velocities in 
extragalactic radio sources which speed up the acceleration
rate.  Using radio interferometry, shock wave velocities
close to
the speed of light have been inferred corresponding 
to typical bulk Lorentz factors in the range
$\Gamma_{\rm jet}=(1-\beta^2)^{-0.5}\sim 2$~to~$10$ ($\beta=u/c$) with
a few cases of still higher values ({\it 10}).
Due to the alignment of the jet axis and the line-of-sight in Mrk~501
superluminal motion has not been observed. 
With $u=c$, $\xi=10$, and taking into account equal
synchrotron and inverse-Compton losses, one obtains 
$E_{\rm max}\sim 4~\Gamma_{\rm jet}(B/{\rm G})^{-0.5} $~TeV from the balance between acceleration gains
and energy losses.  The additional factor $\Gamma_{\rm jet}$ takes into account
the boost in energy due to the relativistic bulk motion.
Therefore, electron maximum energies of $\sim 10$~TeV
as required for Mrk~501 (at least 
5-8~TeV are required for the similar blazar Mrk~421 ({\it 11}))
are
formally allowed, but one certainly has to push the theory to its limits and
this raises a number of
concerns:  (i) the multi-TeV spectrum should show
considerable curvature due to
the (so-called Klein-Nishina) decrease
of the scattering cross section when the energy of the scattered
photon approaches $E_{\rm max}$
and due to the onset of 
electron-positron pair production (the observed multi-TeV spectrum
is consistent with a smooth power law),
(ii) the ratio between the $\gamma$ ray and 
synchrotron (simultaneous) luminosities depends sensitively on 
the jet Lorentz factor $\Gamma_{\rm jet}$ and therefore 
requires  fine-tuning 
(both nearest bright blazars, Mrk~421 and Mrk~501, show
a similar $\gamma$-to-x-ray luminosity ratio), 
(iii)  the magnetic field pressure turns out to be much lower than the 
relativistic electron pressure in the electron acceleration
models which seems inconsistent
with the shock acceleration mechanism (the 
turbulent magnetic field is responsible
for pushing the electrons back and forth across the shock),
larger values of $B$ are also expected from the
adiabatic expansion of a magnetically collimated jet 
(the observed $B$-field
at the tips of the jet in Cygnus~A
is consistent with
adiabatic expansion ({\it 12})), 
and (iv) 
larger values of $\Gamma_{\rm jet}$ could ameliorate the problem that $E_{\rm max}\sim
10$~TeV, however,
one would run into a serious problem with unification models of active
galaxies if $\Gamma_{\rm jet}>10$ would be the rule rather than the
exception ({\it 13}) (number of required host galaxies 
would exceed the number of known radio galaxies).\\

A natural solution to the problem is to assume that the 10~TeV $\gamma$ rays are
due to pion production from accelerated protons.  The balance equation
between energy gains and losses for protons yields maximum energies of
$\sim10^6$~TeV and short variability time scales
$t_{\rm var}\ge 10^5(\xi/\Gamma_{\rm jet})(B/{\rm G})^{-1}$~s
in Mrk~501 ({\it 14}).
The relativistic proton energy loss is dominated by
photo-production of pions in collisions with low-energy
synchrotron photons (originating from accelerated
electrons).  Collisions of the accelerated
protons with matter are negligible due to the low matter
density in relativistic jets (unless a
high-density target moves across the jet ({\it 15})).
The $\gamma$ rays from the decay of the 
neutral pion (far above the observed range of energies)
are subject to pair creation in further collisions with 
the low-energy synchrotron photons ($\gamma+\gamma\rightarrow
e^++e^-$).  This initiates an electromagnetic
cascade shifting the average photon energy to the TeV range and below.
A model based on the combined acceleration of protons
($\gamma$ rays) and electrons (radio-to-x-rays) --
coined the proton blazar model ({\it 16}) --
was fitted to published
data of Mrk~501 in order
to obtain a prediction of its multi-TeV spectrum ({\it 17}).
Data from 1995 and earlier was available for the analysis and covered the
radio-to-x-ray wavelength range including a flux
limit above 100~MeV and an integral flux above 300~GeV
(for details see references in ({\it 17})).
The published flux values showed considerable variability
in the optical-to-x-ray range and
the model spectrum was therefore fitted to match the time-averaged spectrum
(from the fit one obtains $B=37$~G and $\Gamma_{\rm jet}=10$). 
The predicted multi-TeV spectrum is shown in Fig.1 and compared
with the data obtained from recent (1996, 1997) air-Cerenkov observations.
The fairly robust spectral slope of the model spectrum 
fits the
observations very well, 
whereas the absolute flux normalization is somewhat too low.
Considering that the sub-TeV (350~GeV) flux has been reported 
to increase from 1995 to 1997
({\it 9}), the agreement is actually
rather impressive if one scales up the model spectrum accordingly.
Contemporaneous multi-wavelength
observations of blazars such as Mrk~501
will be important to discriminate between
electron-based and proton-based models for the $\gamma$ ray emission
from them.  
Generally, electron-based models require larger values of $\Gamma_{\rm jet}$ and
lower values of $B$ than proton-based models to obtain high-energy
$\gamma$ rays.  \\

There is a further hint that proton acceleration might be important.
Unresolved blazars are the most probable source population to
produce the observed extragalactic $\gamma$ ray background
between 10~MeV and 10~GeV ({\it 18}).
The energy density of this $\gamma$ ray background
is $\simeq 4\times 10^{-6}$~eV~cm$^{-3}$.  
A similar value is found for
an extragalactic flux of protons (with $dN/dE\propto E^{-2}$ 
differential spectrum
between $10^9$~eV and $10^{20}$~eV) providing all of the
observed cosmic rays 
with energies 
above $10^{18.5}$~eV (the so-called ``ankle'' above which the 
very high energy differential cosmic ray
spectrum $\propto E^{-3}$ flattens) ({\it 19}).  
On the assumption that the $\gamma$ rays are from proton ($p$) acceleration,
the comparable energy densities result from simple decay kinematics:
Photo-production of neutral pions ($\pi^0$) ($p+\gamma\rightarrow \pi^\circ+p$)
and their subsequent decay
gives rise to $\gamma$ rays (subject to electromagnetic cascading)
which carry $\sim 1/5$ of the proton energy. 
Charged pions
($p+\gamma\rightarrow \pi^++n$) are produced with 
approximately the same rate
and 
give rise to neutrons ($n$) which carry $\sim
4/5$ of the accelerated proton energy.
Since the neutrons do not scatter off the magnetic field fluctuations
responsible for the acceleration and storage of the charged particles
in the blazar jet,
they must escape the accelerator (energy losses are small
and modify the neutron spectrum only at the very highest energies).
Time dilation allows the most energetic
neutrons
to leave the host galaxy freely before
$\beta$-decay occurs
%Time dilation allows them to leave the host galaxy freely before
%$\beta$-decay occurs
(for protons there would be adiabatic losses due to the magnetic field
in the host galaxy).  
Hence the neutron (and after $\beta$-decay proton) luminosity
is equal to the $\gamma$ ray luminosity within factors of order unity.
An extragalactic origin of the highest energy cosmic rays
is indeed suggested by
the absence of an enhancement
of the cosmic ray flux toward the Galactic disk ({\it 20}) and by the
change in chemical
composition from heavy (protons and ions) to light (protons)
above $10^{18.5}$~eV ({\it 21}). The ultimate challenge to the
hypothesis is the measurement of the high-energy neutrino ($\nu$)
flux associated with the charged pion decay ($\pi^\pm\rightarrow
e^\pm+3\nu$).  The energy density in these multi-TeV neutrinos would be of
the same order of magnitude as that in extragalactic cosmic rays
and $\gamma$ rays. Their measurement therefore constitutes an
experimentum crucis within reach for the planned cubic kilometer
underwater(ice) detectors ({\it 22}).   \\

Although $\gamma$ rays are known from laboratory experiments for their
penetrating power, propagation over intergalactic distances is not without
hurdles.  A diffuse isotropic infrared background (DIRB)
was produced when the first
galaxies formed.  Massive stars in early galaxies produced large amounts
of dust in their winds reprocessing the
visual and ultraviolet light from the stars into infrared light.  
By
colliding with these ample infrared photons, $\gamma$ ray photons
can disappear and turn into electron-positron pairs
({\it 23-25}).  The most numerous infrared
photons above threshold for pair production 
with 10~TeV $\gamma$ rays have wavelengths $\sim 25~\mu$m.
The mean free path ($\lambda_{\gamma\gamma}$) for pair creation at multi-TeV
energies is of the order of the distance ($d$) of Mrk~501.
The exact value depends on the DIRB which is difficult to measure directly owing
to the presence of zodiacal light and galactic cirrus clouds.  \\

One can use the observed power law spectrum ({\it 2}) to put a limit on the
maximum allowed pair attenuation assuming that the observed power law is
the unattenuated spectrum emitted by the source (consistent with
the proton-based model).
In general, only very contrived intrinsic
spectra would look like a smooth power law after the quasi-exponential
attenuation.
The maximum allowed deviation from the power
law $(1-\exp[-d/\lambda_{\gamma\gamma}])$ is taken to be the size of the
statistical error bar at 10~TeV 
yielding an optical depth $\tau_{\gamma\gamma}
=d/\lambda_{\gamma\gamma}<0.7$.  This limit can be relaxed by a factor
not larger than $\sim 2$ admitting for 
weakly absorbed spectra which still approximate a power 
law (see the dashed line in Fig.1).  
There is some dependence of the attenuation on the shape of the 
DIRB spectrum.   Useful models for the spectral shape can be found
in ({\it 23-25}) and yield a similar limit for the $25~\mu$m DIRB
normalization 
$\nu I_\nu(25~\mu{\rm m})<2$~to~$4
$~nW~m$^{-2}$~sr$^{-1}$.
The absence of $\gamma$ ray attenuation in Mrk~501
is consistent with no contribution to the DIRB
other than from the optically selected galaxies
for which one expects $\sim 10\%$ of their optical emission 
to be reprocessed by warm dust yielding
$\nu I_\nu(25\mu{\rm m})\sim 1$~nW~m$^{-2}$~sr$^{-1}$ ({\it 26}), but would
also allow a DIRB stronger by a factor 2 to 4.
A DIRB of at least $\sim 3$~nW~m$^{-2}$~sr$^{-1}$
is suggested by faint infrared galaxy counts and
indicates contributions from 
dust-enshrouded galaxies at redshifts
of $z\sim 3-4$ 
({\it 24}).
Electron-based models for the $\gamma$ ray emission
from Mrk~501 ({\it 9}) predict deviations
from a power law in the multi-TeV range even without
external attenuation and therefore
impose an upper
limit on the DIRB below the lower limit from faint infrared galaxy counts.
If both methods to estimate the DIRB 
(deviations from a power law spectrum in the multi-TeV range,
faint infrared galaxy counts) use
correct assumptions, a cutoff in the $\gamma$ ray
spectrum of Mrk~501 must be present
in the energy range 10 to 30~TeV.
\\

\newpage
\noindent
{\bf References and Notes}
\vskip0.5cm

\noindent\hangindent=0.7cm
1.
M.F. Cawley and T.C. Weekes,  {\em Exp. Astron.} {\bf 6}, 7 (1995).

\noindent\hangindent=0.7cm
2.
F. Aharionan, et al. (HEGRA collaboration),
{\em Astron. Astrophys.} {\bf 327}, L5 (1997).

\noindent\hangindent=0.7cm
3.
C.D. Dermer and R. Schlickeiser, {\em Science} {\bf 257}, 1642 (1992).

\noindent\hangindent=0.7cm
4. 
P.O. Lagage and C.J. Cesarsky, {\em Astron. Astrophys.} {\bf 125}, 249 (1983);
B. Wiebel-Sooth, P.L. Biermann, and H. Meyer, {\em Astron. Astrophys.}, 
in press (1997) (astro-ph/9709253).

\noindent\hangindent=0.7cm
5.
J.R. Jokipii, {\em Astrophys. J.} {\bf 313}, 842 (1987); 
J. Bednartz and M. Ostrowski, {\em
Mon. Not. R. Astr. Soc.} {\bf 283}, 447 (1996).

\noindent\hangindent=0.7cm
6.
S.R. Reynolds,  {\em Astrophys. J.} {\bf 459}, L13 (1996).

\noindent\hangindent=0.7cm
7.
O.C. De Jager, et al., {\em Astrophys. J.}
{\bf 457}, 253 (1996).

\noindent\hangindent=0.7cm
8.
K. Koyama, et al., {\em Nature} {\bf 378},
255 (1995).

\noindent\hangindent=0.7cm
9.
M. Catanese, et al. (Whipple collaboration),
{\em Astrophys. J.} {\bf 487}, L143 (1997);
J. Quinn, et al., {\em Astrophys. J.} {\bf 456},
L83 (1996); E. Pian, et al., {\em Astrophys. J. Lett}, in press (1997)
(astro-ph/9710331).

\noindent\hangindent=0.7cm
10.
G. Ghisellini, P. Padovani, A. Celotti, and L. Maraschi, {\em Astrophys.
J.} {\bf 407}, 65 (1993).

\noindent\hangindent=0.7cm
11.
F. Krennrich, et al. (Whipple collaboration), {\em Astrophys. J.}
{\bf 481}, 758 (1997).

\noindent\hangindent=0.7cm
12.
D.E. Harris, C.L. Carilli, and R.A. Perley, {\em Nature} {\bf 367}, 713 (1994).

\noindent\hangindent=0.7cm
13.
C.M. Urry and P. Padovani, {\em Pub. Astro. Soc. Pac.} {\bf 107}, 803 (1995).

\noindent\hangindent=0.7cm
14.
P.L. Biermann and P.A. Strittmatter, {\em Astrophys. J.} {\bf 322}, 643 (1987).

\noindent\hangindent=0.7cm
15.
A. Dar and A. Laor, {\em Astrophys. J.} {\bf 478}, L5 (1997).

\noindent\hangindent=0.7cm
16.
K. Mannheim, P.L. Biermann, and W.M. Kr\"ulls, {\em Astron. Astrophys.} 
{\bf 251},
723 (1991); K. Mannheim, {\em Astron. Astrophys.} {\bf 269}, 67 (1993).

\noindent\hangindent=0.7cm
17.
K. Mannheim,
S. Westerhoff, H. Meyer, and H.-H. Fink, {\em Astron.
Astrophys.} {\bf 315}, 77 (1996).

\noindent\hangindent=0.7cm
18.
D.A. Kniffen, et al., {\em Astron. Astrophys.
Suppl.} {\bf 120}, 615 (1996); P. Padovani, G. Ghisellini, A.C. Fabian, and
A. Celotti, {\em Mon. Not. R. Astr. Soc.} {\bf 260}, L21 (1993);
C. Impey, {\em Astron. J.} {\bf 112}, 2667 (1996).

\noindent\hangindent=0.7cm
19.
R.J. Protheroe and P.A. Johnson, {\em Astropart. Phys.} {\bf 5}, 215 (1996).

\noindent\hangindent=0.7cm
20.
T. Stanev, P.L. Biermann, J. Lloyd-Evans, J.P. Rachen, and A.A. Watson,
{\em Phys. Rev. Lett.} {\bf 75}, 3065 (1995).

\noindent\hangindent=0.7cm
21.
J.P. Rachen, T. Stanev, and P.L. Biermann, {\em Astron. Astrophys.}
{\bf 273}, 377 (1993); N. Hayashida, et al., {\em Phys. Rev. Lett.} {\bf 77},
1000 (1996).

\noindent\hangindent=0.7cm
22.
T.K. Gaisser, F. Halzen, and T. Stanev,  {\em Phys.~Rep.} {\bf 258}, 173 (1995).

\noindent\hangindent=0.7cm
23.
D. MacMinn and J.R. Primack, {\em Sp. Sci. Rev.} {\bf 75}, 413 (1996).

\noindent\hangindent=0.7cm
24.
A. Franceschini, et al., {\em Astron. Astrophys. Suppl.} {\bf 89}, 285 (1991);
A. Franceschini,  et al., {\em Invited Review
in ESA FIRST symposium (ESA SP 401)}, in press (1997) (astro-ph/9707080);
T. Stanev and A. Franceschini, {\em Astrophys. J. Lett.}, submitted (1997) (astro-ph/9708162).

\noindent\hangindent=0.7cm
25. F.W. Stecker and M.A. Malkan, {\em Astrophys. J.}, in press (1997) (astro-ph/9710072).

\noindent\hangindent=0.7cm
26.
P. Madau, et al., {\em Mon. Not. R. Astr. Soc.} {\bf 283}, 1388 (1996);
B.T. Soifer and G. Neugebauer, {\em Astron. J.} {\bf 101}, 354 (1991).
%R. Abraham, {\em Nature} {\bf 387}, 850 (1997);
%M. Pettini, D.L. King, L.J. Smith, and R.W. Hunstead, {\em Astrophys. J.}
%{\bf 478}, 536 (1997).

\noindent\hangindent=0.7cm
27.
D. Petry, et al., in {\em
Proceedings of the 25th International Cosmic Ray Conference}, 
Durban, South Africa, August 1997, eds. P.A. Evanson et al.,
International Union of Pure and Applied Physics, in press (1998);
S.M. Bradbury, et al. (HEGRA collaboration),
{\it Astron. Astrophys.} {\bf 320}, L5 (1997).

\noindent\hangindent=0.7cm
28.
I thank Peter Biermann, Arnon Dar, John Kirk,
Hinrich Meyer, Joel Primack, Wolfgang Rhode, Frank Rieger,
and the referees for
their critical reading and suggestions for improvements
of the manuscript.  This research was generously supported
by the Deutsche Forschungsgemeinschaft 
under travel grant DFG/Ma 1545/6-1.

\newpage
\noindent
{\bf Figure captions}
\vskip0.5cm

\noindent
{\it Fig.1:}
Differential TeV spectrum of Mrk~501.  
{\em Thin solid line:} proton blazar model prediction based on archival
data from 1995 and earlier ({\it 17}) 
taking into account intergalactic attenuation adopting 
a DIRB spectrum based on a cold+hot dark matter
galaxy formation model from ({\it 23}) and normalized to $\nu
I_\nu (25\mu{\rm m})=1.0$~nW~m$^{-2}$~sr$^{-1}$
(a Hubble
constant of $H_\circ=75$~km~s$^{-1}$~Mpc$^{-1}$ is assumed
throughout the paper).
{\em Thin dashed line:} same without intergalactic attenuation.
{\em Solid thick lines:} model spectrum scaled up by factors 2.5 and 20
corresponding to the increase in the mean sub-TeV flux from 1995 to 1996 and
spring of 1997, respectively ({\it 9}).
{\em Open circles:} HEGRA (High Energy Gamma-Ray Astronomy)
CT1 observation March-August 1996 
(220h) ({\it 27}).
{\em Solid circles:} HEGRA IACT observation March-April 1997 
(26.7h) ({\it 2})

\newpage
\begin{figure*}
\centerline{\psfig{figure=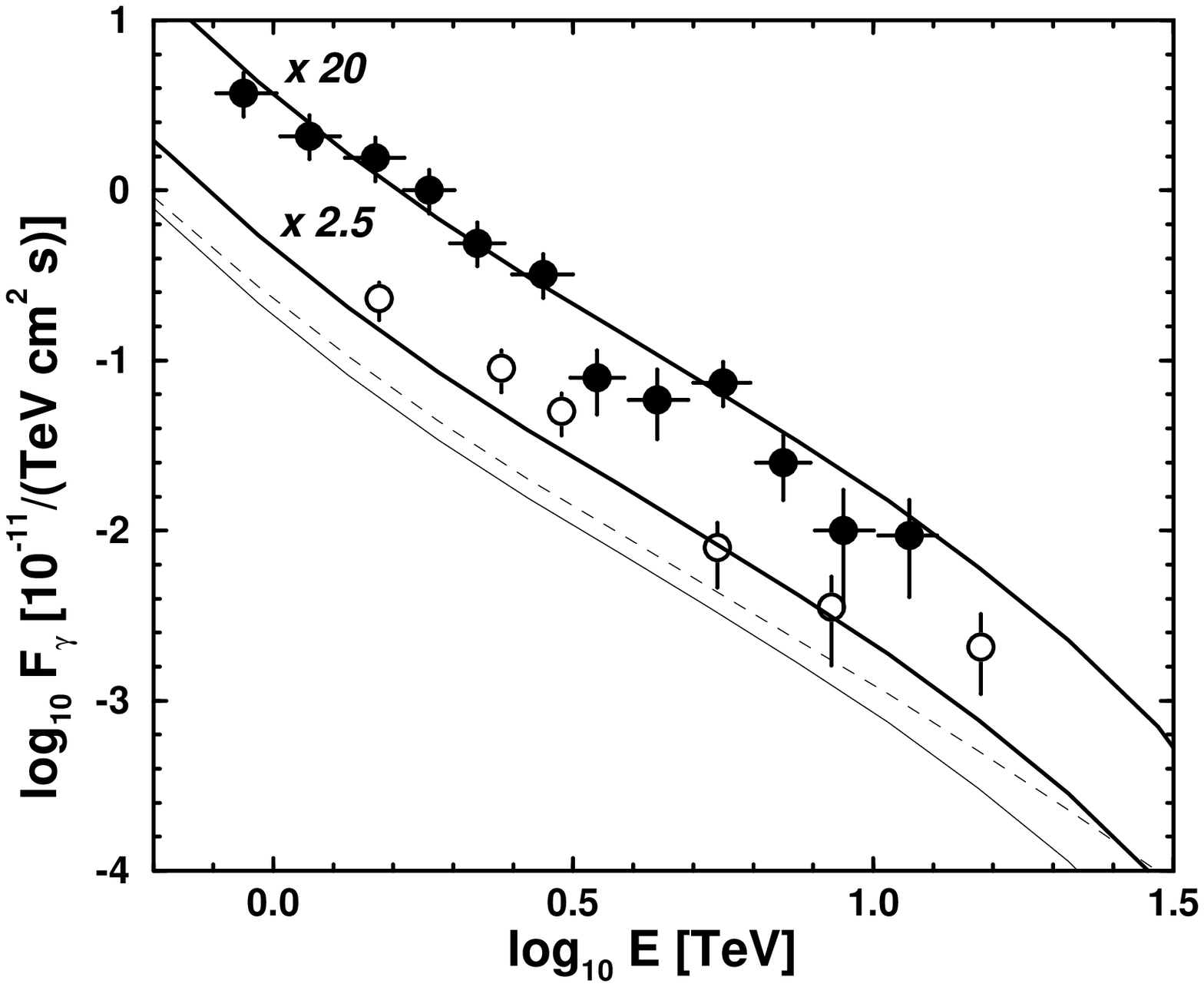,width=14cm}}
\caption[]{ }
\end{figure*}
\end{document}